\documentclass[a4paper,english,conference]{IEEEtran}
\usepackage{latexsym}
\usepackage{float}
\usepackage{amsfonts}
\usepackage{amsbsy}
\usepackage{amssymb}
\usepackage{times}
\usepackage{graphicx}
\usepackage{enumerate}
\usepackage[usenames]{color}
\usepackage[dvips]{pstcol}
\usepackage{epstopdf}
\usepackage{cite}
\usepackage{amssymb}
\usepackage{amsfonts}
\usepackage{graphicx}
\usepackage{epsfig}
\usepackage{psfrag}
\usepackage{xcolor}
\usepackage{amsfonts, bm}
\usepackage{epstopdf}
\usepackage{cite}
\usepackage{color}
\usepackage{xcolor}
\usepackage{subfig}
\usepackage{verbatim}
\usepackage{multirow}
\usepackage{array}
\usepackage{booktabs}
\usepackage{amsthm}

\usepackage[linesnumbered, ruled]{algorithm2e}
\usepackage{algpseudocode}
\usepackage{amsmath}

\IEEEoverridecommandlockouts
\begin{document}
\title{ Robust Semantic Communications Against Semantic Noise }
\author{\IEEEauthorblockN{ Qiyu Hu \IEEEauthorrefmark{1}, Guangyi Zhang \IEEEauthorrefmark{1}, Zhijin Qin \IEEEauthorrefmark{2}, Yunlong Cai \IEEEauthorrefmark{1}, Guanding Yu \IEEEauthorrefmark{1}, and Geoffrey Ye Li \IEEEauthorrefmark{3}  }
\IEEEauthorblockA{\IEEEauthorrefmark{1} College of Information Science and Electronic Engineering, Zhejiang University, Hangzhou, China }
\IEEEauthorblockA{\IEEEauthorrefmark{2} School of Electronic Engineering and Computer Science, Queen Mary University of London, London, UK   }
\IEEEauthorblockA{\IEEEauthorrefmark{3} Department of Electrical and Electronic Engineering, Imperial College London, London, UK \\ E-mail: \{qiyhu, zhangguangyi, ylcai, yuguanding\}@zju.edu.cn, z.qin@qmul.ac.uk, geoffrey.li@imperial.ac.uk} }

\maketitle
\vspace{-3.3em}
\begin{abstract}
Although the semantic communications have exhibited satisfactory performance in a large number of tasks, the impact of semantic noise and the robustness of the systems have not been well investigated. Semantic noise is a particular kind of noise in semantic communication systems, which refers to the misleading between the intended semantic symbols and received ones. 
In this paper, we first propose a framework for the robust end-to-end semantic communication systems to combat the semantic noise. Particularly, we analyze the causes of semantic noise and propose a practical method to generate it. To remove the effect of semantic noise, adversarial training is proposed to incorporate the samples with semantic noise in the training dataset. Then, the masked autoencoder (MAE) is designed as the architecture of a robust semantic communication system, where a portion of the input is masked. To further improve the robustness of semantic communication systems, we firstly employ the vector quantization-variational autoencoder (VQ-VAE) to design a discrete codebook shared by the transmitter and the receiver for encoded feature representation. Thus, the transmitter simply needs to transmit the indices of these features in the codebook.
Simulation results show that our proposed method significantly improves the robustness of semantic communication systems against semantic noise with significant reduction on the transmission overhead. 
\end{abstract}

\IEEEpeerreviewmaketitle

\section{Introduction}
With the development of deep learning (DL) and the increase in deployed devices, more intelligent services have been provided by the networks. These applications generate unprecedented amounts of data for serving different types of tasks, while the conventional communication system if facing the bottleneck to support such massive amount of data \cite{IoTData}. To address this issue, semantic communication emerged as a key technology and have received great attention \cite{Principle}. Different from conventional communications, only essential semantic information relevant to the task is extracted from source message and transmitted to the receiver, which further compresses the data while reserving the task-related information \cite{JSCC}. The existing works on semantic communications can be categorized into: (i) data reconstruction, where the global semantic information behind data is extracted and the data is reconstructed based on the received semantic information \cite{JSCC,DeepSC,JSCCf}; (ii) task execution, where only the task-related semantic information is encoded at the transmitter and then directly applied for task execution at the receiver \cite{TransIoT,ImagRetri}.

Though the aforementioned DL-based semantic communication systems have exhibited very good performance in certain tasks, the impact of semantic noise and the system robustness have not been considered. There exists a particular type of semantic noise in semantic communication systems that has not been well studied \cite{Principle}. Semantic noise is a kind of noise that causes misunderstanding of semantic information and decoding errors. It results in the misleading between the transmitted semantic symbols and the received ones, which can be introduced in the stages of semantic encoding, data transmission, and decoding \cite{SemanMaga}. 
Particularly, the causes of semantic noise for different sources, e.g., text and image, are different. The semantic noise in text refers to semantic ambiguity, where small changes to words, e.g., synonym replacement, may make the DL model misunderstand the semantic meaning of the sentence. In this paper, we focus on the semantic noise in the image domain, which can be modeled based on the adversarial samples. In particular, some subtle modifications can be added to images that are barely noticeable to humans. Adversarial samples mislead DL models and cause significant performance degradation, but if observed by humans, they look identical to the original images. A number of methods have been proposed for generating adversarial samples in the image domain, such as the fast gradient sign method (FGSM) \cite{FGSM} and the projected gradient descent (PGD) \cite{PGD}.
The existing DL-based systems are vulnerable and particularly unstable to these adversarial samples, where small and imperceptible perturbations of the data samples are sufficient to fool them and result in incorrect results \cite{intriguing}. To improve the robustness of DL model against adversarial samples, defensive distillation \cite{distillation} and adversarial training \cite{WeightPerturb} have been developed. 

Although there have been lots of methods for generating adversarial samples in the field of image processing, the semantic noise model in wireless communication area has not been well investigated, by leveraging the wireless environment. Moreover, the performance of these methods against adversarial samples is unsatisfactory and even deteriorates in the original samples. The complexity of these methods is high and the training process usually takes long to converge. In addition, they do not consider the impact of the wireless channels and the transmission overhead in communications. Therefore, it is important to model the semantic noise in communication field and design a robust semantic communication system that can effectively combat the impacts of semantic noise with lower transmission overhead. 

\begin{figure*}[t]
\begin{centering}
\includegraphics[width=0.85\textwidth]{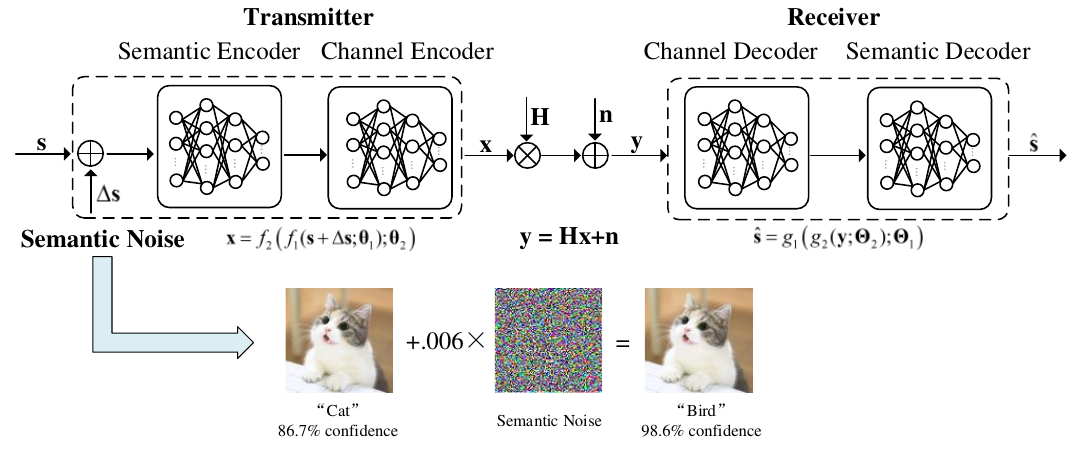}
\par\end{centering}
\caption{The framework of semantic communication system with semantic noise.}
\label{FrameArchitect}
\end{figure*}

In this paper, we propose a DL-enabled end-to-end robust semantic communication system to combat the semantic noise and our main contributions are summarized as follows.

\begin{itemize}
\item We firstly model the semantic noise by employing an iterative FGSM to instantly generate the semantic noise. To combat the semantic noise, we propose an adversarial training method with weight perturbation, which incorporates the samples with semantic noise in the training dataset to solve a complicated min-max optimization problem. 

\item Then, the masked autoencoder (MAE) with Transformer blocks \cite{MAE} is designed as the architecture of robust semantic communication systems, where a portion of the original image is masked and only the unmasked portion is encoded and transmitted. 

\item We are the first to employ the vector quantization-variational autoencoder (VQ-VAE) to design a discrete codebook shared by the transmitter and the receiver for encoded feature representation, which improves the system robustness. Since some operations are non-differentiable, we employ the straight-through estimator \cite{VQVAE} to train the codebook. 
The transmitter simply needs to send the indices of the features in the codebook to the receiver, which significantly reduces the transmission overhead.  

\item Simulation results show that our proposed method significantly improves the robustness of semantic communication systems against semantic noise with significant reduction on the transmission overhead.  
\end{itemize}


\section{Framework of Robust Semantic Communications} \label{System}
In this section, we propose the framework of robust end-to-end semantic communication systems to deal with the semantic noise. 

\subsection{Semantic Communication Systems}
As shown in Fig. \ref{FrameArchitect}, the transmitter maps the source, $\mathbf{s}$, into a symbol stream, $\mathbf{x}$, and then passes it through the physical channel with transmission impairments. The received symbol stream, $\mathbf{y}$, is decoded at the receiver to have an estimation of the source, $\hat{\mathbf{s}}$. 
Both the transmitter and the receiver are represented by deep neural networks (DNNs). In particular, the DNNs at the transmitter consist of the semantic encoder and channel encoder, while the DNNs at the receiver consist of the semantic decoder and channel decoder. The semantic encoder learns to transform the transmitted data into an encoded feature vector while the semantic decoder learns to recover the transmitted data from the received signals. Moreover, the channel encoder and channel decoder aim at eliminating the signal distortion caused by the wireless channel. 

We consider a system with $N_{t}$ transmit antennas and $N_{r}$ receive antennas. The encoded symbol stream can be represented by
\begin{equation} \label{SemCom01}
\mathbf{x}=f_{2}\big( f_{1}(\mathbf{s};\bm{\theta}_{1} ); \bm{\theta}_{2}  \big),
\end{equation}
where $\mathbf{x}\in \mathbb{C}^{N_{t}\times 1}$, $\bm{\theta}_{1}$ and $\bm{\theta}_{2}$ denote the trainable parameters of the semantic encoder, $f_{1}(\cdot)$, and the channel encoder, $f_{2}(\cdot)$, respectively. Subsequently, the signal received at the receiver, $\mathbf{y}\in \mathbb{C}^{N_{r}\times 1}$, is given by
\begin{equation} \label{SemChannel}
\mathbf{y}=\mathbf{H}\mathbf{x}+\mathbf{n},
\end{equation} 
where $\mathbf{H}\in \mathbb{C}^{N_{r}\times N_{t}}$ denotes the channel matrix and $\mathbf{n}\sim \mathcal{CN}(\bm{0}, \sigma^{2}\mathbf{I})$ is the additive white Gaussian noise (AWGN). Correspondingly, the decoded signal is given as
\begin{equation} \label{SemCom02}
\hat{\mathbf{s}}=g_{1}\big( g_{2}(\mathbf{y};\bm{\Theta}_{2} ); \bm{\Theta}_{1}  \big),
\end{equation}
where $\bm{\Theta}_{1}$ and $\bm{\Theta}_{2}$ denote the trainable parameters of the semantic decoder, $g_{1}(\cdot)$, and the channel decoder, $g_{2}(\cdot)$, respectively. For clarity, we denote $\bm{\theta}$ as the set of trainable parameters and $f_{\bm{\theta}}(\cdot)$ as the DNNs in the considered semantic communication systems. Thus, we have $\hat{\mathbf{s}}=f_{\bm{\theta}}(\mathbf{s})$. 

\subsection{Generation of Semantic Noise} 
As shown in Fig. \ref{FrameArchitect}, the semantic noise, $\Delta \mathbf{s}$, is added at the source. Considering the scenario where a malicious attacker downloads the image dataset, adds semantic noise to each image, and then uploads the modified dataset. The semantic noise will mislead the DL models to generate wrong results or decision makings. However, since the semantic noise is so subtle that the legitimate users can hardly notice, they will use these contaminated images as usual. Moreover, semantic noise also exists in nature, and images obtained by taking pictures of the adversarial samples could also cause misclassification \cite{Physical}.

The goal of the semantic communication system is to minimize the loss function for serving a specific task, e.g., the cross entropy for classification task. In contrast, the semantic noise aims to maximize the loss function. Let $\mathcal{S} = \{ \mathbf{s}_{1}, \mathbf{s}_{2}, \cdots, \mathbf{s}_{I} \}$ be a set of images sampled from the training dataset. Then, the generation of semantic noise for the $i$-th image $\mathbf{s}_{i}$ can be modeled as solving the following optimization problem 
\begin{subequations}  \label{NoiseProblem}
\begin{eqnarray}
\textrm{P1}: & \max\limits_{\Delta \mathbf{s}_{i}} & \mathcal{L}(f_{\bm{\theta}} (\mathbf{s}_{i} + \Delta \mathbf{s}_{i}), \mathbf{z}_{i}) \\
& \textrm{s.t.} & \|\Delta \mathbf{s}_{i} \|_{p} \leq \epsilon, \label{PowerCons}
\end{eqnarray}
\end{subequations}
where $\mathbf{s}_{i}$ and $f_{\bm{\theta}} (\mathbf{s}_{i} + \Delta \mathbf{s}_{i})$ denote the $i$-th input image and output of the DNN, respectively, $\Delta \mathbf{s}_{i}$ is the semantic noise generated for the $i$-th image, and $\mathcal{L}(\cdot)$ denotes the loss function of the DNN for a specific task. In addition, $\mathbf{z}_{i}$ denotes the target associated with $\mathbf{s}_{i}$, e.g., the true label for classification task. Note that $\|\cdot \|_{p}$ is the $p$-norm and the constraint \eqref{PowerCons} limits the power of semantic noise to avoid being observed by human. Unless otherwise stated, we select $p=2$ in this paper.

To solve this problem, we employ the FGSM \cite{FGSM}, which linearizes the loss function as
\begin{equation}
\mathcal{L}(f_{\bm{\theta}} (\mathbf{s}_{i} + \Delta \mathbf{s}_{i}), \mathbf{z}_{i})  \! \approx \!  \mathcal{L}(f_{\bm{\theta}} (\mathbf{s}_{i}), \mathbf{z}_{i}) + (\Delta \mathbf{s}_{i})^{T} \nabla_{\mathbf{s}_{i}} \mathcal{L}( f_{\bm{\theta}} (\mathbf{s}_{i}), \mathbf{z}_{i}).
\end{equation}
It is minimized by setting $\Delta \mathbf{s}_{i} = -\alpha \nabla_{\mathbf{s}_{i}} \mathcal{L}( f_{\bm{\theta}} (\mathbf{s}_{i}), \mathbf{z}_{i})$,
where $\alpha$ is a scaling factor to constrain the power of semantic noise to $\epsilon$ in \eqref{PowerCons}. Then, we obtain the optimal semantic noise with power $\epsilon$ as 
\begin{equation}
\Delta \mathbf{s}_{i} = \epsilon \textrm{sign}\big( \nabla_{\mathbf{s}_{i}} \mathcal{L}( f_{\bm{\theta}} (\mathbf{s}_{i}), \mathbf{z}_{i}) \big),
\end{equation}
where $\textrm{sign}(x)=1$ for $x\geq 0$ and $\textrm{sign}(x)=-1$ for $x<0$. Then, the contaminated sample with semantic noise becomes $\mathbf{s}'_{i}=\mathbf{s}_{i}+\Delta \mathbf{s}_{i}$. To increase the impact of $\Delta \mathbf{s}_{i}$ on the system, we modify the FGSM into an iterative process,
\begin{equation} \label{IteraFGSM}
\mathbf{s}_{i}'^{(k+1)} = \Pi_{\epsilon} \big( \mathbf{s}_{i}'^{(k)} + \alpha \cdot \textrm{sign}( \nabla_{\mathbf{s}_{i}} \mathcal{L}( f_{\bm{\theta}} (\mathbf{s}_{i}'^{(k)}), \mathbf{z}_{i}) ) \big),
\end{equation}
where $k$ denotes the iteration index and $\Pi$ is the projection operator. We select $\alpha$ satisfying $K\alpha >\epsilon$ to ensure that we can take full advantage of the noise power $\epsilon$, where $K$ denotes the number of iterations.

\subsection{Adversarial Training} 
The key idea of adversarial training against semantic noise is to add the samples corrupted by the semantic noise into the training dataset. In particular, the trainable parameters, $\bm{\theta}$, and semantic noise, $\Delta \mathbf{s}_{i}$, are updated iteratively to improve the model robustness. It can be formulated as solving the following min-max optimization problem
\begin{subequations}  \label{AdvTrain}
\begin{eqnarray}
\textrm{P2}: & \min\limits_{\bm{\theta}} & \frac{1}{I}\sum\limits_{i=1}^{I} \max_{\Delta \mathbf{s}_{i}} \mathcal{L}( f_{\bm{\theta}} (\mathbf{s}_{i}+\Delta \mathbf{s}_{i}), \mathbf{z}_{i} ) \\
& \textrm{s.t.} & \|\Delta \mathbf{s}_{i}\|_{p}\leq \epsilon,
\end{eqnarray}
\end{subequations}
where $I$ denotes the number of training samples. 
To solve (P2), the following two steps are executed iteratively: (i) Compute $\mathbf{s}'_{i}$ based on \eqref{IteraFGSM} and the semantic noise is obtained by $\Delta \mathbf{s}_{i}= \mathbf{s}'_{i}-\mathbf{s}_{i}$, which aims at maximizing the loss function. Note that $\bm{\theta}$ is fixed in this step and we add the samples $\mathbf{s}'_{i}$ into the training dataset; (ii) Update $\bm{\theta}$ by stochastic gradient descent (SGD) based on the training samples $\mathbf{s}'_{i}$ to minimize the loss function.

To further improve the robustness against semantic noise, we propose to add the weight perturbation $\bm{\nu}$ on trainable parameters and reformulate the problem as
\begin{subequations}  \label{ATAWP}
\begin{eqnarray}
\textrm{P3}: &\min\limits_{\bm{\theta}} \max\limits_{\bm{\nu}} & \frac{1}{I}\sum\limits_{i=1}^{I} \max\limits_{\Delta \mathbf{s}_{i}} \mathcal{L}( f_{\bm{\theta}+\bm{\nu}} (\mathbf{s}_{i}+\Delta \mathbf{s}_{i}), \mathbf{z}_{i} ) \\
& \textrm{s.t.} & \|\Delta \mathbf{s}_{i}\|_{p}\leq \epsilon, \quad \|\bm{\nu} \|_{p}\leq \gamma \|\bm{\theta} \|_{p}. 
\end{eqnarray}
\end{subequations}
Intuitively, the input perturbation (semantic noise $\Delta \mathbf{s}_{i}$) and weight perturbation $\bm{\nu}$ lead to the increase of loss function $\mathcal{L}(\cdot)$ for the $i$-th sample and all the samples, respectively. Thus, the two ``max" makes the inner maximization problem solve better, which results in a better solution of the whole min-max problem \cite{WeightPerturb}.
To solve (P3), we conduct the following three steps iteratively: 
\begin{itemize}
\item Compute $\mathbf{s}'_{i}$ based on \eqref{IteraFGSM} and the semantic noise is obtained by $\Delta \mathbf{s}_{i}= \mathbf{s}'_{i}-\mathbf{s}_{i}$, which aims to maximize the loss function. Note that $\bm{\theta}$ and $\bm{\nu}$ are fixed in this step and the computed samples $\mathbf{s}'_{i}$ are added into the training dataset. 

\item Update $\bm{\nu}$ to maximize the loss function by using one step of forward and backward propagation with fixed $\bm{\theta}$ and $\Delta \mathbf{s}_{i}$.

\item Update $\bm{\theta}$ to minimize the loss function by SGD with fixed $\bm{\nu}$ based on the training samples $\mathbf{s}'_{i}$.
\end{itemize}

\section{Masked Autoencoder with Discrete Codebook}  \label{Architect}
In this section, we propose an architecture of the robust semantic communication systems with MAE and discrete codebook.

\begin{figure*}[t]
\begin{centering}
\includegraphics[width=0.88\textwidth]{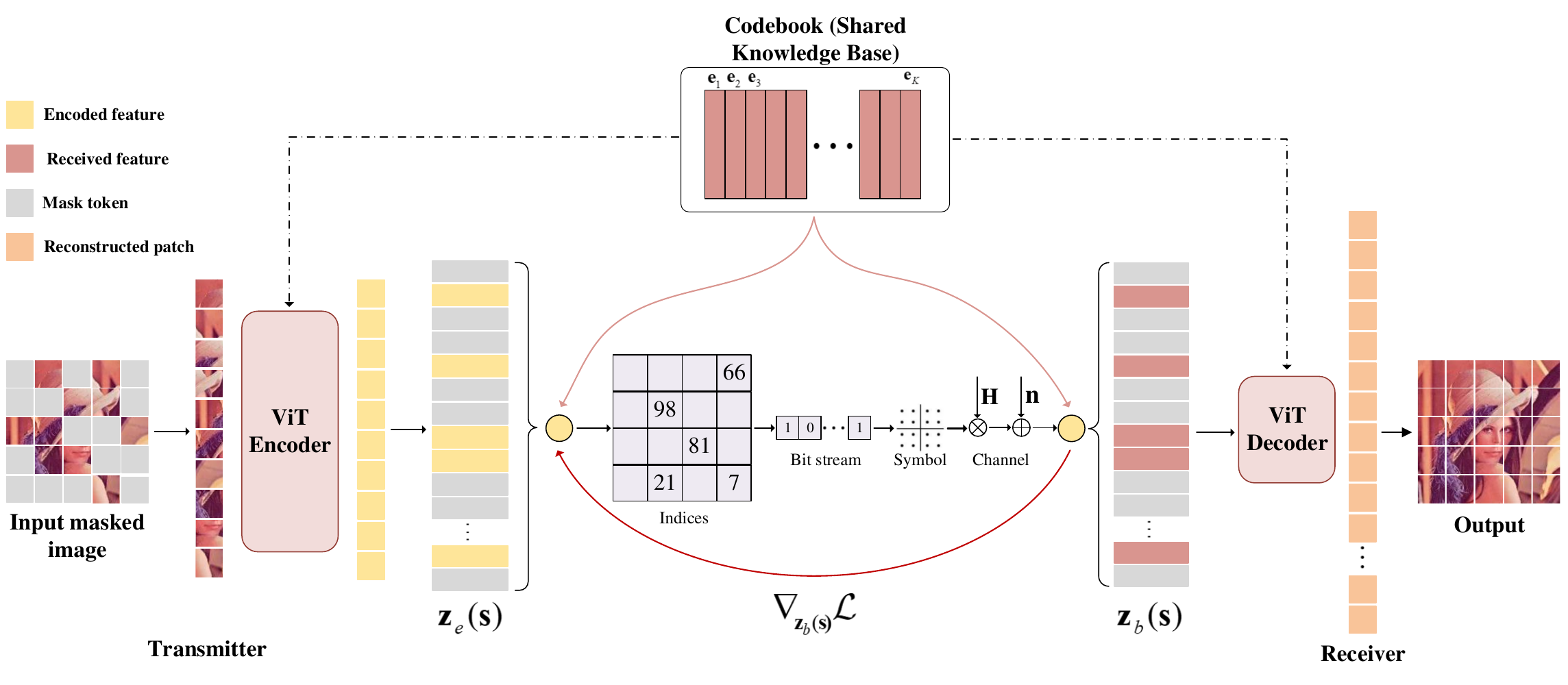}
\par\end{centering}
\caption{The architecture of MAE with discrete feature representation.}
\label{MAEStruc}
\end{figure*}

\subsection{Architecture of MAE} 
There exists information redundancy in various kinds of sources and the image has heavy spatial redundancy. A missing patch in the image can be recovered from neighboring patches with the understanding of parts, objects, and scenes. Thus, the strategy of randomly masking partial patches is an efficient approach to create a challenging task that requires the model to build a comprehensive understanding of image statistics and semantic information, which also reduces the information redundancy. Moreover, since the semantic noise is added in the patches of the image, the masking operation can eliminate the effects of semantic noise to some extent.

As presented in Fig. \ref{MAEStruc}, we employ the MAE with vision Transformer (ViT) architecture, where we randomly mask patches from input images and aim to reconstruct the missing patches \cite{MAE}. The MAE is one kind of autoencoder that reconstructs the original image from partial observations. Unlike conventional autoencoders, we adopt an asymmetric encoder-decoder architecture.
In particular, the encoder only needs to process a small portion of unmasked patches and maps them to the encoded features for transmission, which significantly reduces the training time and memory consumption. It removes the masked patches and embeds the unmasked patches with their positional information in the original image, and then processes them via a series of Transformer blocks \cite{MAE}. 
In contrast, the input to the decoder is the full set of tokens consisting of: (i) encoded features of unmasked patches; and (ii) mask tokens, as shown in Fig. \ref{MAEStruc}. Each mask token is a shared and learned vector that indicates the presence of a missing patch to be predicted. We add positional embeddings to all tokens in this full set. Without this, mask tokens would have no information about their locations in the image. 
Moreover, the decoder is only used during pre-training to perform the image reconstruction task, while the encoder is employed to extract features of input images. Thus, the decoder architecture can be flexibly designed that is independent of the encoder design. Compared with the conventional autoencoder in communication systems, the MAE is equipped with following advantages:
\begin{itemize} 
\item With such an asymmetrical design, the encoder only processes the unmasked patches and the lightweight decoder reconstructs the image from the encoded features and mask tokens, which significantly reduces computational complexity and training time. 

\item The pre-trained MAE can be employed for different downstream tasks, e.g., classification, by simply changing the structure of lightweight decoder and fine-tuning the MAE within a short time. 

\item Transmitting the encoded features of unmasked patches to the decoder at the receiver leads to a large reduction in transmission overhead. 

\item The masking operation can reduce the impacts of semantic noise.
\end{itemize}

\subsection{Discrete Codebook for Encoded Feature Representation} 
We aim to employ the VQ-VAE to design a discrete codebook for the encoded feature space and represent these encoded features by the basis vectors in this codebook. It considers important task-related features and neglects task-unrelated features with noise and imperceptible details. Specifically, we set these basis vectors as trainable parameters and train them together with the parameters of encoder and decoder. The encoder network outputs continuous encoded features and then maps them into the discrete indices of basis vectors in the trained codebook. This design comes with the following advantages:
\begin{itemize} 
\item It is simple to train the codebook and it does not suffer from large variance, which makes the semantic communication system more stable. 

\item It can mitigate the impact of semantic noise.

\item The transmitter simply needs to send the indices of basis vectors, which significantly reduces the transmission overhead.
\end{itemize}

\subsubsection{Codebook Design}
As presented in Fig. \ref{MAEStruc}, we define the codebook of encoded features as $\mathcal{E}\triangleq \big[\mathbf{e}_{1}, \mathbf{e}_{2}, \cdots, \mathbf{e}_{J} \big] \in \mathbb{R}^{J\times D}$, which consists of $J$ basis vectors $\{\mathbf{e}_{j} \in \mathbb{R}^{D}, j\in 1, 2, \cdots, J\}$, and $D$ is the dimension of each basis vector $\mathbf{e}_{j}$. The model takes an input $\mathbf{s}$ and it passes through an encoder to produce the encoded feature vector $\mathbf{z}_{e}(\mathbf{s})$. Then, it is mapped to a basis vector $\mathbf{z}_{b}(\mathbf{s})$ by a nearest neighbor look-up
\begin{equation} \label{VQVAEform}
\mathbf{z}_{b}(\mathbf{s})=\textrm{arg} \min\limits_{ \mathbf{e}_{j} } \big\| \mathbf{z}_{e}(\mathbf{s}) - \mathbf{e}_{j} \big\|_{2}, \forall \mathbf{e}_{j}.
\end{equation}
Then, $\mathbf{z}_{b}(\mathbf{s})$ is input to the decoder. We can treat this forward computation as a layer of DNN with a particular non-linear function that maps the encoded feature vector $\mathbf{z}_{e}(\mathbf{s})$ to a basis vector $\mathbf{z}_{b}(\mathbf{s})$. The basis vectors $\{\mathbf{e}_{j}, \forall j\}$ in the codebook $\mathcal{E}$ are trained together with the parameters of encoder and decoder. 
However, the operation \eqref{VQVAEform} is non-differentiable. Thus, in back propagation, we approximate the gradient by straight-through estimator and copy gradients from decoder input $\mathbf{z}_{b}(\mathbf{s})$ to encoder output $\mathbf{z}_{e}(\mathbf{s})$ \cite{VQVAE}. Therefore, the nearest basis vector $\mathbf{z}_{b}(\mathbf{s})$ is passed to the decoder in forward propagation, and during the back propagation, the gradient $\nabla_{ \mathbf{z}_{b}(\mathbf{s}) } \mathcal{L}_{c}$ is passed unaltered to the encoder. Note that the output of encoder and the input of decoder have the same dimension $D$ and the gradients contain useful information for guiding the encoder to update its parameters to minimize the loss function $\mathcal{L}_{c}$. 

We design the loss function consisting of three components, which are employed to train
different parts of parameters 
\begin{equation} \label{VQVAEloss}
\! \mathcal{L}_{c}(\! \mathbf{s}, \mathbf{z}; \bm{\theta}, \mathbf{e}_{j} \!) \!\! = \!\! \big\|  \hat{\mathbf{s}}-\mathbf{z} \big\|_{2} \!+\! \big\| \! \textrm{ng}\big[ \! \mathbf{z}_{e}(\mathbf{s}) \! \big] - \mathbf{e}_{j} \! \big\|_{2} \!+\! \beta \big\| \mathbf{z}_{e}(\mathbf{s}) - \textrm{ng}\big[ \! \mathbf{e}_{j} \big] \! \big\|_{2},
\end{equation}
where $\mathbf{s}$, $\hat{\mathbf{s}}$, and $\mathbf{z}$ denote the input, output, and true label of the network, respectively, $\bm{\theta}$ denotes the trainable parameters of the original DNN, and $\beta$ is the hyper-parameter. The symbol ``ng" represents that there is no gradient passed to its operand and its gradient is zero. It effectively constrains its operand to be a non-updated constant. The first term is the reconstruction loss that trains the parameters of encoder and decoder. Due to the straight-through gradient estimation of mapping from $\mathbf{z}_{e}(\mathbf{s})$ to $\mathbf{z}_{b}(\mathbf{s})$, the basis vectors $\{\mathbf{e}_{j}, \forall j\}$ receive no gradient from the reconstruction loss $\|\hat{\mathbf{s}}-\mathbf{z}\|_{2}$. Therefore, in order to train the basis vectors $\{\mathbf{e}_{j}, \forall j\}$, we propose to employ the $l_{2}$ error to move the basis vectors towards the encoded features $\mathbf{z}_{e}(\mathbf{s})$, as shown in the second term of \eqref{VQVAEloss}. Since the volume of the encoded feature space is dimensionless, the codebook $\mathcal{E}$ can grow arbitrarily and causes the training process to diverge if the basis vectors $\{\mathbf{e}_{j}, \forall j\}$ are not trained as fast as the encoder parameters. To address this issue, we add the third term in \eqref{VQVAEloss}. In summary, the decoder is optimized by the first loss term only, the encoder is optimized by the first and the last loss terms, and the basis vectors are optimized by the middle loss term. 
 
\subsubsection{Efficient Transmission}
We assume that the transmitter and receiver share the codebook $\mathcal{E}$ consisting of the basis vectors $\{\mathbf{e}_{j}, \forall j\}$, which are fixed after the training stage. Thus, for each encoded feature output by the encoder, the transmitter simply needs to send the index of the corresponding basis vector, which significantly reduces the transmission overhead.

\subsubsection{Robust Codebook Against Semantic Noise}

\begin{figure}[t]
\begin{centering}
\includegraphics[width=0.35\textwidth]{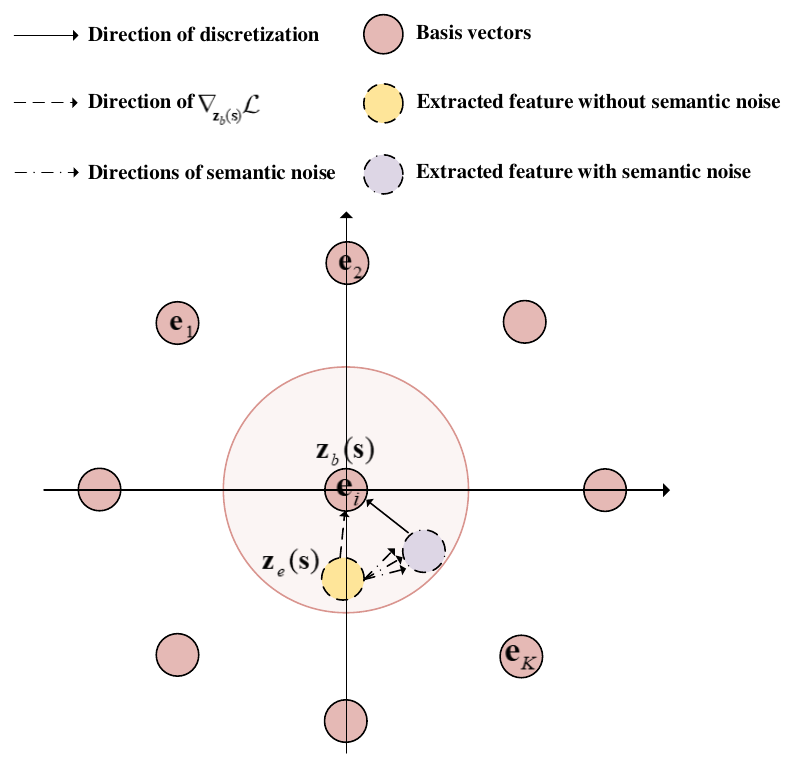}
\par\end{centering}
\caption{The discrete codebook against semantic noise. }
\label{VQVAENoise}
\end{figure}

The discrete representation is efficient to reduce the impacts of semantic noise. As shown in Fig. \ref{VQVAENoise}, the semantic noise causes the extracted feature to move towards some certain directions. Only if it does not leave far away from the basis vector corresponding to the original extracted feature that is not affected by semantic noise, the impact of semantic noise can be eliminated by this discrete representation.

\section{Simulation Results} \label{Simulation}
In this section, we compare the proposed MAE with conventional source coding and channel coding approaches under the AWGN, Rayleigh, and Rician channels. We adopt CIFAR 10 as the dataset, which consists of $60,000$ images. The average size of joint photographic experts group (JEPG) images in the dataset is $5,108$ bytes and the number of patches of each image is $14\times 14$. The masking ratio of MAE is $0.5$ and the codebook size is $256$ which requires $8$ bits for transmitting each index. We adopt the 16-QAM for modulation and the code rate of low-density parity-check code (LDPC) is selected as $1/2$.
The number of iterations to generate semantic noise is set as $K=5$ and its power is $\epsilon=0.012$. The number of layers and multi-heads of MAE encoder are set as $14$ and $12$, respectively. The decoder consists of $3$ fully-connected layers. For different downstream tasks, e.g., classification, we simply need to change the dimension of its last layer and fine-tune the pre-trained MAE.
We consider the image classification task and provide the performance of the following methods:
\begin{itemize}
\item MAE+SN: The proposed MAE with the semantic noise.

\item MAE+SN without AT: The proposed MAE with the semantic noise but without the adversarial training. 

\item MAE: The proposed MAE without the semantic noise.

\item JPEG+LDPC+SN: The conventional scheme that adopts JPEG for the image source coding, LDPC for the channel coding, and the ViT as classifier with the semantic noise and adversarial training.

\item JPEG+LDPC: The conventional scheme without the semantic noise.

\item ViT+SN: The conventional ViT with the semantic noise, which does not employ the adversarial training, mask technique, and discrete codebook. We assume the error-free transmission without the impact of channel impairments.

\item ViT: The conventional ViT without the semantic noise under the error-free transmission.
\end{itemize}

\begin{table}
\centering  
\caption{The number of transmitted symbols for one image.}    
\label{Overhead} 
\begin{tabular}{|c|c|c|c|}
\hline
Methods & JPEG+LDPC &  Proposed MAE & Ratio     \\ \hline
The number of symbols  & $20432$ & $196$ & $0.95\%$    \\ \hline
\end{tabular}
\end{table}

Table \ref{Overhead} presents the transmission overhead of: (i) The conventional JPEG+LDPC: $5108 \times 8 \times 2 \div 4=20432$ symbols/image; (ii) the proposed MAE with masking technique and discrete codebook: $14\times 14\times 0.5\times 8\div 4=196$ symbols/image.

\begin{figure}[t]
\begin{centering}
\includegraphics[width=0.39\textwidth]{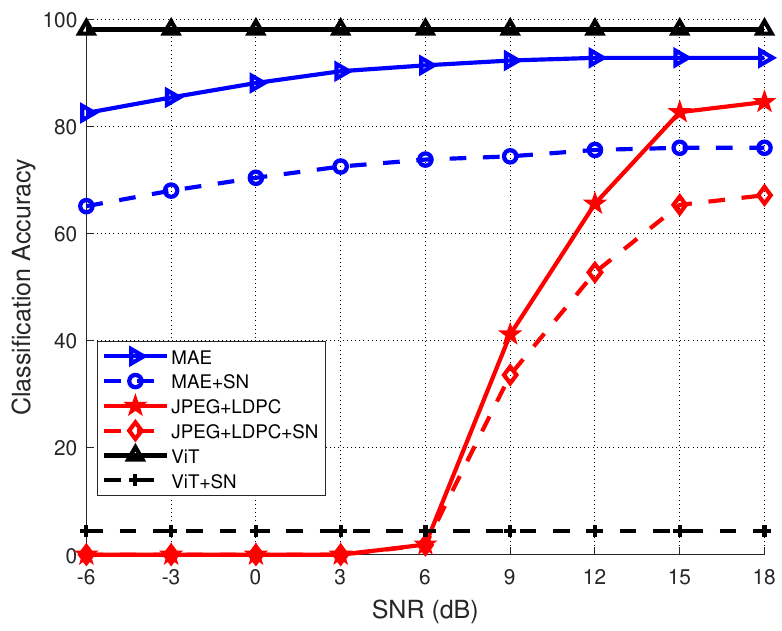}
\par\end{centering}
\caption{The classification accuracy versus SNR.}
\label{AccSNR}
\end{figure}

Fig. \ref{AccSNR} shows the classification accuracy versus SNR with Rayleigh channel. We train the proposed MAE model with SNR$=0$ dB and test it in SNR from $-6$ dB to $18$ dB. It is readily seen that the classification accuracy achieved by the proposed MAE model and the conventional schemes increases with SNR. The proposed MAE significantly outperforms the conventional schemes, especially in low SNR. It is because that the bit error rate (BER) is high in low SNR and the proposed MAE is more robust by transmitting the indices of extracted task-related features in the trained codebook. Furthermore, MAE+SN significantly outperforms ViT+SN and MAE approaches ViT, which demonstrates that the proposed MAE with adversarial training, mask technique, and discrete codebook can improve the system robustness by reducing the impacts of semantic noise and channel. 

\begin{figure}[t]
\begin{centering}
\includegraphics[width=0.4\textwidth]{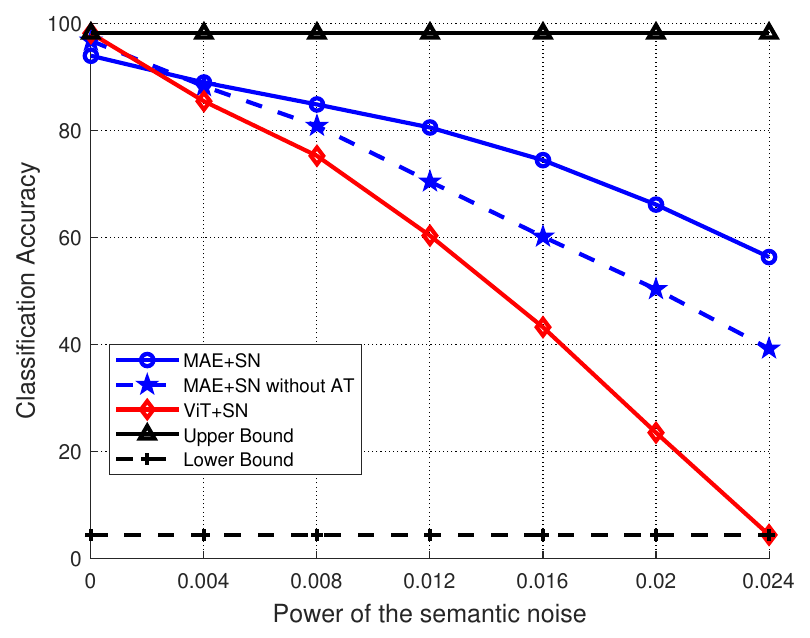}
\par\end{centering}
\caption{The classification accuracy versus the power of semantic noise.}
\label{SemPower}
\end{figure}

Fig. \ref{SemPower} illustrates the classification accuracy versus the power of semantic noise $\epsilon$. The upper bound and lower bound are achieved by ViT+SN without semantic noise and maximum power of semantic noise, respectively. It is readily seen that the classification accuracy achieved by all the schemes decreases with $\epsilon$. The proposed MAE+SN significantly outperforms the MAE+SN without AT and ViT+SN, especially when $\epsilon$ is large. It shows the superiority of the proposed MAE with adversarial training, mask technique, and discrete codebook against semantic noise.

\begin{figure}[t]
\begin{centering}
\includegraphics[width=0.39\textwidth]{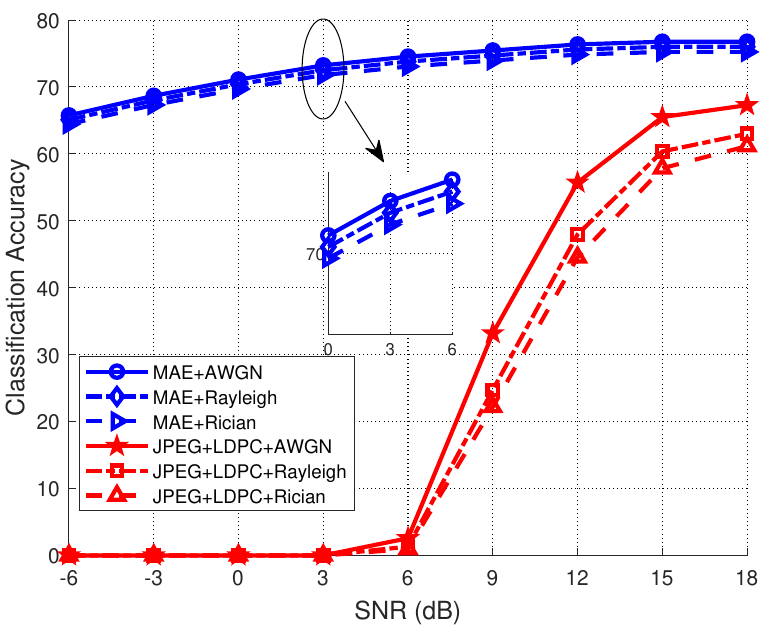}
\par\end{centering}
\caption{The generalization ability for different channels with semantic noise.}
\label{AccChannel}
\end{figure}

Fig. \ref{AccChannel} presents the generalization ability for different channels. We train the proposed MAE model in AWGN channel with SNR$=0$ dB and test it in AWGN, Rayleigh, and Rician channels for SNR from $-6$ dB to $18$ dB with the semantic noise. It can be seen that the proposed model achieves similar classification accuracy in different channels, which significantly outperforms the conventional schemes, especially at low SNR. Moreover, the classification accuracy of conventional scheme degrades in Rayleigh and Rician channels. It shows that our proposed model can adapt to different channels with good performance.   

\section{Conclusion} \label{Conclusion}
In this paper, we first proposed a framework for the robust semantic communication system with the semantic noise. Particularly, the semantic noise has been modeled and a practical method has been proposed to generate it. To combat the semantic noise, the adversarial training that incorporates the samples with the semantic noise in the training dataset has been developed. Then, the MAE with masking strategy was designed as the architecture of the robust semantic communication system. To further improve the robustness of the system, we designed a discrete codebook shared by the transmitter and the receiver for the encoded feature representation. Simulation results show that our proposed method can significantly improve the robustness of semantic communication systems against the semantic noise with significant reduction on transmission overhead.

\bibliographystyle{IEEEtran}
\bibliography{IEEEabrv,Semantic}

\begin{thebibliography}{10}
\providecommand{\url}[1]{#1}
\csname url@samestyle\endcsname
\providecommand{\newblock}{\relax}
\providecommand{\bibinfo}[2]{#2}
\providecommand{\BIBentrySTDinterwordspacing}{\spaceskip=0pt\relax}
\providecommand{\BIBentryALTinterwordstretchfactor}{4}
\providecommand{\BIBentryALTinterwordspacing}{\spaceskip=\fontdimen2\font plus
\BIBentryALTinterwordstretchfactor\fontdimen3\font minus
  \fontdimen4\font\relax}
\providecommand{\BIBforeignlanguage}[2]{{%
\expandafter\ifx\csname l@#1\endcsname\relax
\typeout{** WARNING: IEEEtran.bst: No hyphenation pattern has been}%
\typeout{** loaded for the language `#1'. Using the pattern for}%
\typeout{** the default language instead.}%
\else
\language=\csname l@#1\endcsname
\fi
#2}}
\providecommand{\BIBdecl}{\relax}
\BIBdecl

\bibitem{IoTData}
M.~Mohammadi, A.~Al-Fuqaha, S.~Sorour, and M.~Guizani, ``Deep learning for
  {IoT} big data and streaming analytics: A survey,'' \emph{IEEE Commun.
  Surveys Tuts.}, vol.~20, no.~4, pp. 2923--2960, Jun. 2018.

\bibitem{Principle}
Z.~Qin, X.~Tao, J.~Lu, and G.~Y. Li, ``Semantic communications: Principles and
  challenges,'' \emph{arXiv preprint arXiv:2201.01389}, 2021.

\bibitem{JSCC}
E.~Bourtsoulatze, D.~Burth~Kurka, and D.~Gunduz, ``Deep joint source-channel
  coding for wireless image transmission,'' \emph{IEEE Trans. Cognit. Comm.
  Netw.}, vol.~5, no.~3, pp. 567--579, Sep. 2019.

\bibitem{DeepSC}
H.~Xie, Z.~Qin, G.~Y. Li, and B.-H. Juang, ``Deep learning enabled semantic
  communication systems,'' \emph{IEEE Trans. Signal Process.}, vol.~69, pp.
  2663--2675, Apr. 2021.

\bibitem{JSCCf}
D.~B. Kurka and D.~Gunduz, ``{DeepJSCC}-f: Deep joint source-channel coding of
  images with feedback,'' \emph{IEEE J. Select. Areas Inf. Theory}, vol.~1,
  no.~1, pp. 178--193, May 2020.

\bibitem{TransIoT}
C.-H. Lee, J.-W. Lin, P.-H. Chen, and Y.-C. Chang, ``Deep learning-constructed
  joint transmission-recognition for {Internet of Things},'' \emph{IEEE
  Access}, vol.~7, pp. 76\,547--76\,561, Jun. 2019.

\bibitem{ImagRetri}
M.~Jankowski, D.~Gunduz, and K.~Mikolajczyk, ``Wireless image retrieval at the
  edge,'' \emph{IEEE J. Select. Areas Commun.}, vol.~39, no.~1, pp. 89--100,
  Jan. 2021.

\bibitem{SemanMaga}
G.~Shi, Y.~Xiao, Y.~Li, and X.~Xie, ``From semantic communication to
  semantic-aware networking: Model, architecture, and open problems,''
  \emph{IEEE Commun. Mag.}, vol.~59, no.~8, pp. 44--50, Aug. 2021.

\bibitem{FGSM}
I.~J. Goodfellow, J.~Shlens, and C.~Szegedy, ``Explaining and harnessing
  adversarial examples,'' in \emph{Proc. Int'l. Conf. Learn. Represent.
  (ICLR)}, May 2015, pp. 1--11.

\bibitem{PGD}
A.~Madry, A.~Makelov, L.~Schmidt, D.~Tsipras, and A.~Vladu, ``Towards deep
  learning models resistant to adversarial attacks,'' in \emph{Proc. Int'l.
  Conf. Learn. Represent. (ICLR)}, May 2018, pp. 1--10.

\bibitem{intriguing}
C.~Szegedy, W.~Zaremba, I.~Sutskever, J.~Bruna, D.~Erhan, I.~Goodfellow, and
  R.~Fergus, ``Intriguing properties of neural networks,'' in \emph{Proc.
  Int'l. Conf. Learn. Represent. (ICLR)}, Apr. 2014, pp. 1--10.

\bibitem{distillation}
N.~Papernot, P.~McDaniel, X.~Wu, S.~Jha, and A.~Swami, ``Distillation as a
  defense to adversarial perturbations against deep neural networks,'' in
  \emph{IEEE Symp. Secur. Privacy (SP)}, 2016, pp. 582--597.

\bibitem{WeightPerturb}
D.~Wu, S.-T. Xia, and Y.~Wang, ``Adversarial weight perturbation helps robust
  generalization,'' in \emph{Proc. Adv. Neural Inf. Process. Syst. (NIPS)},
  Dec. 2020, pp. 1--20.

\bibitem{MAE}
K.~He, X.~Chen, S.~Xie, Y.~Li, P.~Dollar, and R.~Girshick, ``Masked
  autoencoders are scalable vision learners,'' \emph{arXiv preprint
  arXiv:2111.06377}, 2021.

\bibitem{VQVAE}
A.~van~den Oord, O.~Vinyals, and K.~Kavukcuoglu, ``Neural discrete
  representation learning,'' in \emph{Proc. Adv. Neural Inf. Process. Syst.
  (NIPS)}, Dec. 2017, pp. 6309--6318.

\bibitem{Physical}
A.~Kurakin, I.~Goodfellow, and S.~Bengio, ``Adversarial examples in the
  physical world,'' in \emph{Proc. Int'l. Conf. Learn. Represent. Workshop
  (ICLR)}, May 2017, pp. 1--13.

\end{thebibliography}

\end{document}